\documentclass[conference]{IEEEtran}
\IEEEoverridecommandlockouts

\usepackage{cite}
\usepackage{amsmath,amssymb,amsfonts}
\usepackage{algorithm}
\usepackage{algorithmicx}
\usepackage{graphicx}
\usepackage[noend]{algpseudocode}
\usepackage{textcomp}
\usepackage{xcolor}
\def\BibTeX{{\rm B\kern-.05em{\sc i\kern-.025em b}\kern-.08em
    T\kern-.1667em\lower.7ex\hbox{E}\kern-.125emX}}
\usepackage{hyperref}
\hypersetup{
    colorlinks=true,
    linkcolor=black,
    filecolor=black,
    anchorcolor = black,
    citecolor=black,
    urlcolor=blue
    }
    
\begin{document}

\title{
Accelerating Deep Reinforcement Learning for Digital Twin Network Optimization with Evolutionary Strategies}

\author{\IEEEauthorblockN{Carlos~Güemes-Palau\IEEEauthorrefmark{1}, Paul~Almasan\IEEEauthorrefmark{1}, Shihan~Xiao\IEEEauthorrefmark{2}, Xiangle~Cheng\IEEEauthorrefmark{2}, Xiang~Shi\IEEEauthorrefmark{2}\\ Pere~Barlet-Ros\IEEEauthorrefmark{1}, Albert~Cabellos-Aparicio\IEEEauthorrefmark{1}}
\IEEEauthorblockA{\IEEEauthorrefmark{1}Barcelona Neural Networking Center, 
Universitat Politècnica de Catalunya, Spain\\
}
\IEEEauthorblockA{\IEEEauthorrefmark{2}Network Technology Lab., Huawei Technologies Co.,Ltd.\\}}

\maketitle

\begin{abstract}
The recent growth of emergent network applications (e.g., satellite networks, vehicular networks) is increasing the complexity of managing modern communication networks. As a result, the community proposed the Digital Twin Networks (DTN) as a key enabler of efficient network management. Network operators can leverage the DTN to perform different optimization tasks (e.g., Traffic Engineering, Network Planning). 

Deep Reinforcement Learning (DRL) showed a high performance when applied to solve network optimization problems. In the context of DTN, DRL can be leveraged to solve optimization problems without directly impacting the real-world network behavior. However, DRL scales poorly with the problem size and complexity. In this paper, we explore the use of Evolutionary Strategies (ES) to train DRL agents for solving a routing optimization problem. The experimental results show that ES achieved a training time speed-up of 128 and 6 for the NSFNET and GEANT2 topologies respectively. 
\end{abstract}


\section{Introduction}

In the last years, different industry sectors adapted the digital twin paradigm to model complex systems. A digital twin can be seen as a digital representation of a real-world system or phenomena. In a networking context, the community proposed the Digital Twin Network (DTN) as a key technology to enable efficient operation in modern networks \cite{zhou-nmrg-digitaltwin-network-concepts-06, almasan2022digital}. The DTN is a digital representation of the physical network that enables operators to design high performance optimization solutions for complex scenarios, to efficiently plan the next network upgrades or to perform troubleshooting, among others \cite{almasan2022digital}. 

Modern networks have seen a growing trend in network traffic, number of connected devices and a proliferation of emergent network applications (e.g., vehicular networks, IoT) \cite{6964248}. These trends made modern networks become highly dynamic and heterogeneous, increasing the complexity of efficient network management. In addition, network applications impose different requirements on the real-world network (e.g., low latency, high throughput). However, these applications share the underlying network's limited resources, making it difficult for the network operator to fulfill different requirements at the same time. 

The DTN is a key enabler for efficient optimization in modern networks\cite{almasan2022digital}. In particular, the DTN models the real-world network dynamics such as the growing trend in network traffic or users. This enables the network operator to manage the network taking into account the underlying network events. For example, the network operator can efficiently plan the network or perform traffic engineering considering the network's traffic trends. In other words, the DTN created new sprouts of network applications that they don't only take into account the current network state, but also the past and the future.

Network optimization consists in operating the network with the goal of using the network resources efficiently. In the DTN context, the optimization process consists in the interaction between a network optimizer and the DTN. The role of the DTN is to estimate the real-world network performance metrics (e.g., end-to-end delay). These metrics are used to guide the network optimizer towards high quality solutions (e.g., routing configurations, scheduling policies). 

Deep Reinforcement Learning (DRL) is a technology that showed high performance when applied on solving network optimization problems \cite{10.1145/3452296.3472902, almasan2021enero, 9651930, 10.1145/3098822.3098843}. DRL differs from traditional optimization solutions (e.g., ILP, CP) because it leverages the \textit{knowledge} learned in past optimizations, which is beneficial for optimizing in highly dynamic scenarios. In addition, DRL enables high performance and real-time network operation in large network optimization scenarios (e.g., large network topologies) \cite{almasan2021enero}.

The main issue with DRL-based solutions is that they scale poorly. The DRL agent learns from a sequential interaction with an environment. When DRL is applied to large optimization scenarios (e.g., large network topologies) or highly complex optimization problems (e.g., Traffic Engineering), the interaction with the environment becomes computationally intensive. For example, we implemented the DRL solution meant for solving network routing optimization from \cite{almasan2019deep} using the PPO algorithm\cite{schulman2017proximal}. The DRL agent took 3 minutes to train in the 14-node NSFNET topology\cite{hei2004wavelength}, but when trained on the 24-node GEANT2 topology \cite{barreto2012fast} it took nearly an hour. Hence, a simple exponential regression estimates that to train in a 300-node topology would take around two weeks. This illustrates how applying DRL for solving real-world problems is limited by its sequential nature.

In this paper, we explore an alternative method to train DRL agents using Evolution Strategies (ES)\cite{salimans2017evolution}. Specifically, ES removes the traditional backpropagation from the DRL agent's training process and applies a highly parallelizable weight optimization algorithm to find the best set of weights of a Neural Network (NN). In addition, we integrated a Graph Neural Network (GNN) in the DRL agent to enable efficient optimization on graphs (i.e., network topology). 

To showcase the capabilities of combining ES with DRL on graph-based optimization problems, we apply it on a routing optimization scenario in Optical Transport Networks (OTN). Specifically, we implemented the DQN algorithm from \cite{almasan2019deep} using PPO, which corresponds to our computational time baseline. We applied ES to the PPO algorithm version and executed it using up to 64 processes. 

\section{Problem Statement}

\begin{figure}[t]
    \centering
    \includegraphics[width=0.8\linewidth]{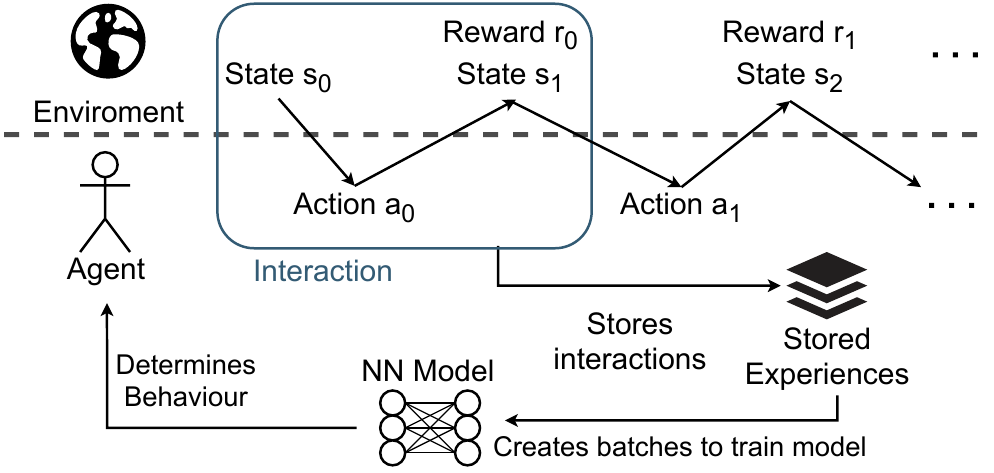}
    \vspace{-2mm}
    \caption{General overview of the DRL agent training process. The interactions between the agent and the environment are stored for training the weights of the NN later on. These weights determine the agent's behavior.}
    \label{fig:rlseq}
\vspace{-4mm}
\end{figure}

In the DRL context, the training process requires a DRL agent and an environment. The DRL agent acts on the environment, changing its internal state. The environment defines the problem at hand that needs to be solved. After each action, the environment returns a new state and a reward, indicating the DRL agent if the action performed was good or bad. 

Generally, the training process is composed by two main phases. First, the DRL agent interacts with the environment and collects all the states and rewards returned by the environment. These interactions are stored and later on are used by different DRL algorithms to make the agent converge to a good policy\cite{schulman2017proximal, lillicrap2015continuous}. Second, the stored experiences are used to train the NN weights following a supervised learning fashion. Traditionally, the training process of DRL agents follows a sequential pattern that alternates between these two phases, which is illustrated in  Figure~\ref{fig:rlseq}.

The NN weights of the DRL agent are typically trained using gradient descent with a backpropagation algorithm. Therefore, the second phase can be easily accelerated using GPUs. However, to accelerate the first phase is challenging. Specifically, the sequential nature of the interaction between the DRL agent and the environment cannot be broken. In addition, the first phase scales poorly with the problem size. This is because when the optimization problem becomes larger or grows in complexity, it typically requires larger interactions with the environment. 

\section{Network optimization problem}
\label{sec:network_problem}

To showcase the capabilities of applying ES to accelerate the training of DRL agents, we train a DRL agent in a routing optimization use case. Particularly, we consider a SDN-based scenario where the DRL agent is placed in the control plane and is provided with a global view of the real-world Optical Transport Network (OTN)~\cite{strand2001issues}. The goal of the DRL agent is to optimize the routing configuration by efficiently using the network's resources.

The optimization problem can be described as a classical resource allocation problem\cite{almasan2019deep,chen2018deep,suarez2019routing}. It is defined by an OTN whose links have a limited bandwidth capacity and a set of traffic demands, defined by the tuple \textit{\{src, dst, bandwidth\}}. These demands need to be routed through a particular sequence of links that connect the source and the destination nodes. Once routed, the traffic demand occupies the links' utilization, and the resources are not freed until the end of a DRL episode. Therefore, the problem consists of allocating the maximum traffic volume in the network. A traffic demand cannot be allocated when the bandwidth is larger than the available link capacity for any possible path, indicating the end of a DRL episode. 

\section{Proposed Solution}

\begin{figure}[t]
    \centering
    \includegraphics[width=0.8\linewidth]{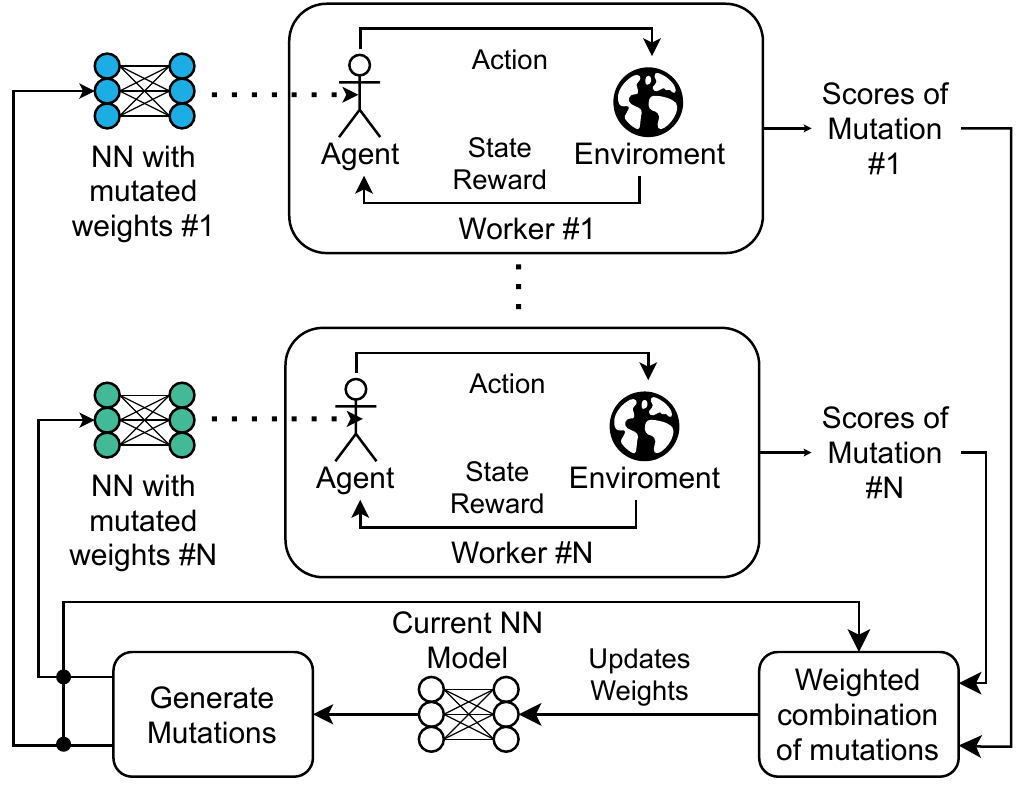}
    \vspace{-2mm}
    \caption{General overview of ES applied to DRL.}
    \label{fig:es}
\vspace{-4mm}
\end{figure}

In this paper, we adapt Evolutionary Strategies (ES) \cite{salimans2017evolution} to accelerate the training of a DRL agent that performs routing optimization. ES are a set of optimization algorithms that are part of the family of evolutionary algorithms \cite{salimans2017evolution}. Specifically, ES follow an iterative method that tries to converge to a set of NN weights that maximizes some objective function.

At the beginning of the optimization process, the objective function and a set of initial NN weights $\theta_{init}$ are given. Then, these weights are perturbed, creating multiple mutations of the original weights. Each mutation is then evaluated and has associated a score, indicating how good or bad are the weights to satisfy the objective function. Afterwards, the mutations are combined into one update using natural gradient ascent. The optimization process starts again using the new weights $\theta_{new}$, and continues until convergence. Intuitively, the weights are updated to become similar to those of the better performing mutations. By the end of the ES algorithm, the NN weights $\theta_{final}$ will converge at a point where no mutations can be generated that obtain higher scores.

\subsection{Design}

In a DRL context, the initial weights for the DRL agent policy are obtained using some well-established method \cite{glorot2010understanding}. The weight perturbations are created using noise sampled from a multivariate Gaussian distribution. The number of mutations is a hyperparameter that has a trade-off between the quality of exploration (i.e., large number of mutations) and speed (i.e., low number of mutations). Each mutation defines a new DRL agent policy that is evaluated by making it interact with the environment. The resulting sum of rewards from each mutation is used to obtain the evaluation score. Figure~\ref{fig:es} shows a general overview of ES applied to DRL.

\renewcommand\algorithmicindent{.5em}

\begin{algorithm}[t!]
\caption{Improved Parallelized NES}\label{alg:INES}
\begin{algorithmic}[1]
\small{\Require Learning rate $\alpha$,  noise standard deviation for mutations $\sigma$, initial policy parameters $\theta_0$, fitness function $F$, number of mutations $k$
\For{each worker $i = 1,...,n$}
    \State Obtain assigned number of mutations $k_i$
    \If{worker $i$ is coordinator} \Comment{Coordinator thread}
        \For{$t=0,1,2,...$}
            \State Sample global perturbations $\epsilon_1, ..., \epsilon_k; \epsilon \sim \mathcal{N}(0,I)$
            \State Compute local returns $F' \gets F(\theta_t + \sigma\epsilon_i), \epsilon_i \in k_i$
            \State Receive and aggregate returns other workers: $\hat F$
            \State Obtain solution update $\Delta\theta \gets \alpha \frac{1}{n\sigma}\sum_{j=1}^{n}{\hat F_j\epsilon_j}$
            
            \State Send $\Delta\theta$ to all worker threads
            \State Update solution $\theta_{t+1} \gets \theta_t + \Delta\theta$
        \EndFor
    \Else \Comment{Worker thread}
        \For{$t=0,1,2,...$}
            \State Sample local perturbations $\epsilon_1, \epsilon_2, ..., \epsilon_{k_i}, \epsilon_j \sim \mathcal{N}(0,I)$
            \State Compute returns $F' \gets F(\theta_t + \sigma\epsilon_i), \epsilon_i \in k_i$
            \State Send returns $F'$ to coordinator thread
            \State Receive $\Delta\theta$ and Update Solution $\theta_{t+1} \gets \theta_t + \Delta\theta$
        \EndFor
    \EndIf
\EndFor}
\end{algorithmic}
\end{algorithm}

The particularities of our graph-based optimization problem forced us to make some adaptations to the original ES method \cite{salimans2017evolution}. Originally, the algorithm was totally distributed, where each worker updates its weights independently. 
Instead, we designed a centralized version, where only one of the workers performs the weight updates and propagates the main NN weights to the other workers. We refer to this worker as the \textit{coordinator}. 

With the centralized method it is no longer necessary for the perturbations to be replicated across all workers. This drastically reduces the memory cost. In our work, the workers only communicate with the coordinator, which results in less messages being sent.

We also include some of the optional improvements proposed by the original paper, including fitness shaping, mirrored sampling and added noise to the action probabilities distribution of the agent \cite{salimans2017evolution}.

Algorithm~\ref{alg:INES} describes the training process of a DRL agent using ES. The role of the workers is to sample the perturbations, evaluate the mutations, send the returns to the coordinator and update the solution. The coordinator has a more complex task which starts by generating the perturbations. Once these have been evaluated by the workers, the coordinator retrieves the scores and computes the new NN weights. This new set of weights is then sent to all the workers.

\section{Experimental Results}

In this section we train a DRL agent using ES to efficiently allocate traffic demands in the OTN routing scenario described in Section~\ref{sec:network_problem}. Specifically, we adapt the code from an existing solution and implement it so we can train DRL agents with PPO\cite{almasan2019deep} and ES. We use the PPO implementation as the baseline. Specifically, we train two different agents on the 14-node NSFNET \cite{hei2004wavelength} and the 24-node GEANT2 topologies \cite{barreto2012fast}. We also train a third agent over both topologies simultaneously, so it learns to optimize both at the same time. Then compare the training times of the agents when using PPO and ES. We also perform preliminary experiments to find the proper hyperparameters such as number of perturbations or standard deviation of the mutations.

We execute the DRL agent with ES with varying number of workers according to the hardware resources available. First, we execute ES in a local server with 16 CPU cores with support for 32 concurrent threads of execution, where we use ES with 1, 2, 4, 8, 16 and 32 workers. Second, we execute ES on an Amazon Web Services (AWS) c5a.16xlarge instance with 64 CPU cores, where we use ES with 16, 32 and 64 workers.

\begin{figure}[t!]
    \centering
    \includegraphics[width=0.85\linewidth]{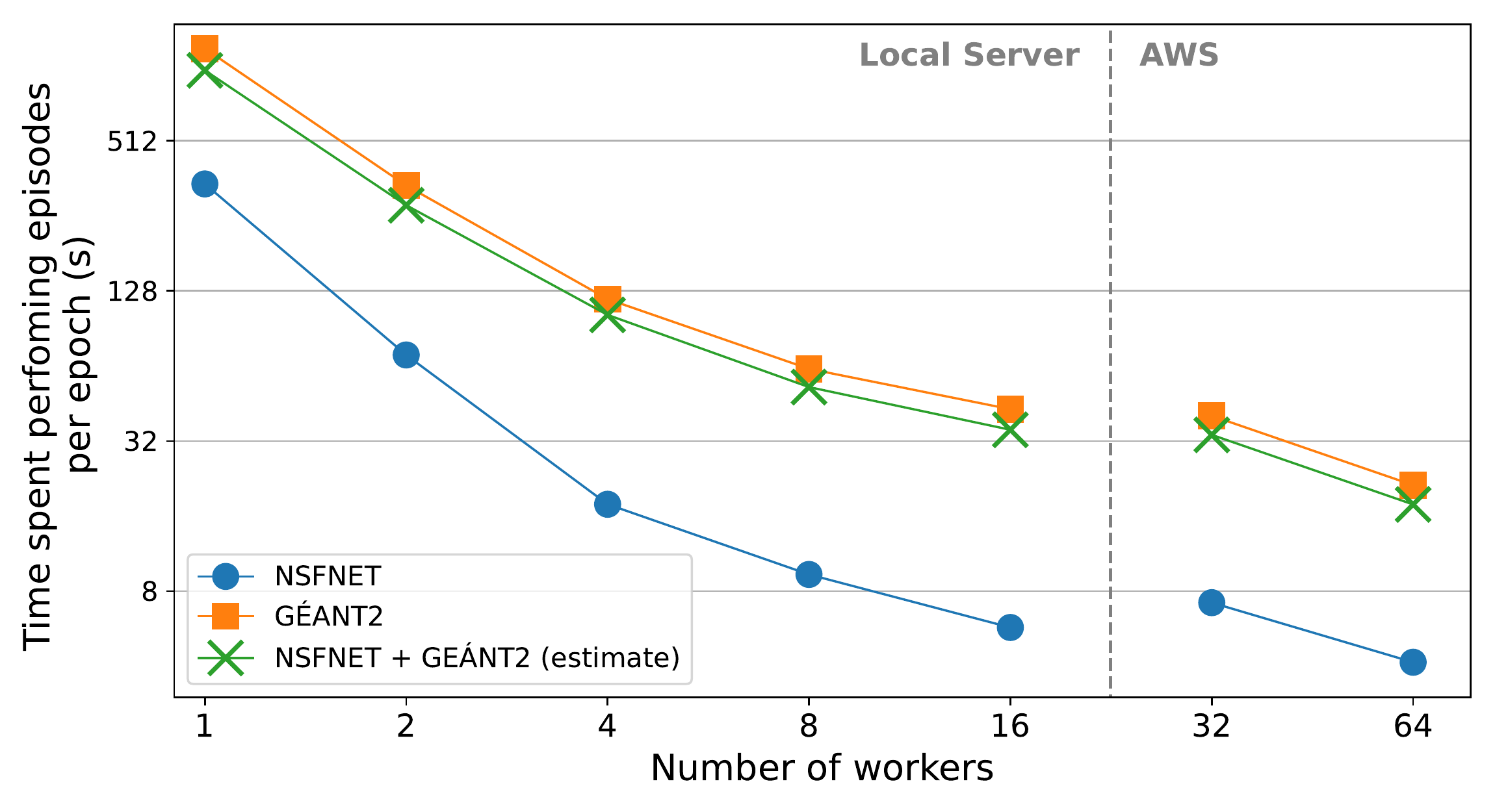}
    \vspace{-2mm}
    \caption{Impact of the number of workers on the time spent per epoch performing episodes. Notice that the y-axis is in logarithmic scale.}
    \label{fig:res_1}
\vspace{-4mm}
\end{figure}

\begin{figure}[t!]
    \centering
    \includegraphics[width=0.85\linewidth]{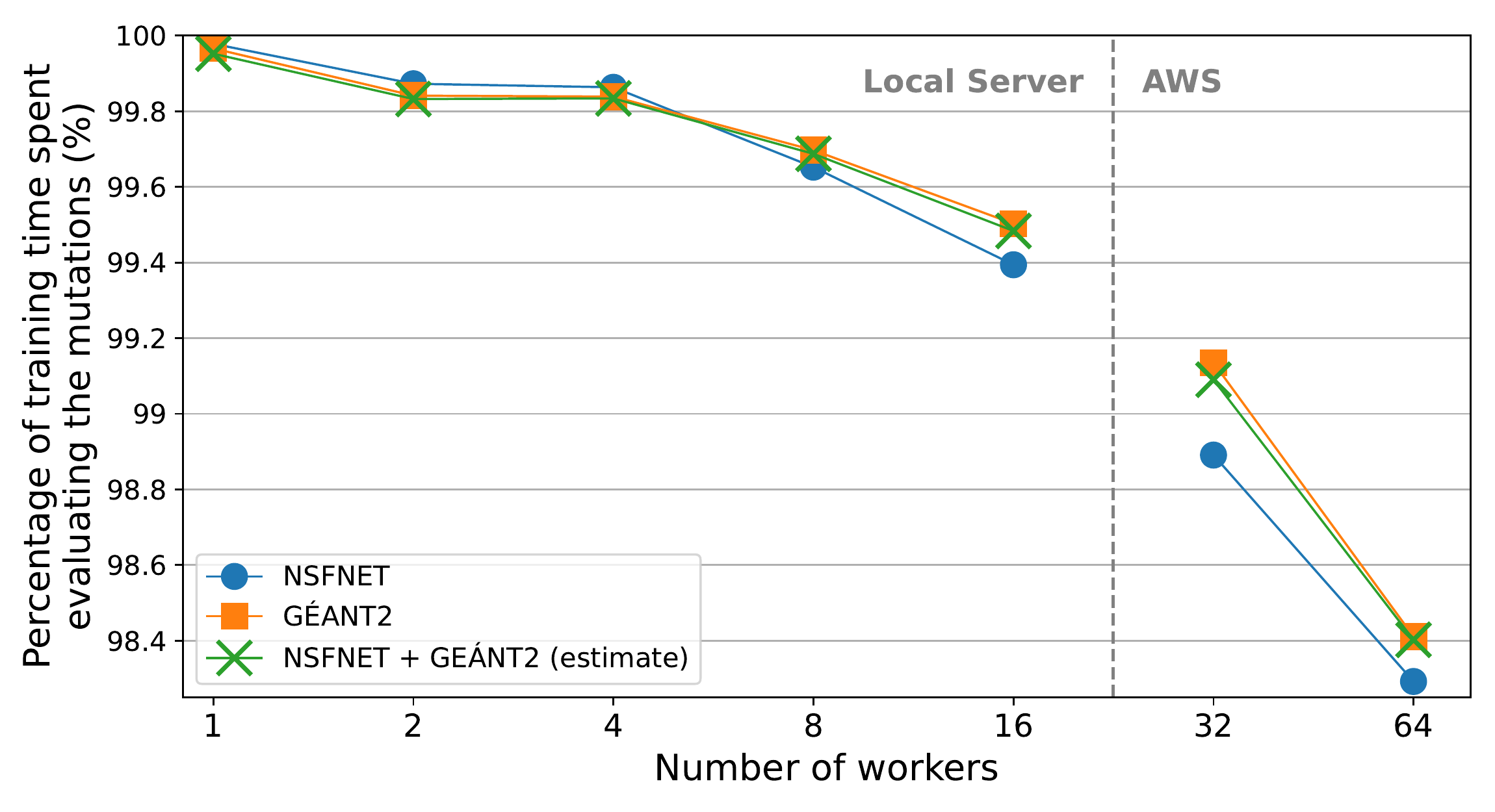}
    \vspace{-2mm}
    \caption{Impact of the number of workers over the relative time spent per epoch performing episodes.}
    \label{fig:res_2}
\end{figure}

The source code is written in Python, with the usage of libraries such as NumPy for vector computation \cite{numpy}, Tensorflow for representing the GNN \cite{tensorflow}, NetworkX for representing the network state \cite{networkx}, the OpenAI Gym framework for representing the DRL environment \cite{brockman2016openai}, and MPI4PY for establishing the communication between workers\cite{mpi4py}. The source code, together with all the training and evaluation results presented in this paper, are publicly available\footnote{\url{https://github.com/BNN-UPC/DRL-ES-OTN}}.

\subsection{Training time speed-up}

In this experiment we want to see the impact of the number of workers in the training time of the DRL agent with ES. Figure~\ref{fig:res_1} shows the experimental results based on the time spent per epoch to evaluate all the mutations across all configurations. The experimental results show that an increasing the number of workers results in a linear decrease in time spent interacting with the environment, resulting in lower training times. 

Figure~\ref{fig:res_2} shows the relative amount of training time dedicated solely to interact with the environment. In other words, it indicates the percentage of time dedicated to evaluate the mutations. The results indicate that increasing the number of workers decreased the amount of time spent interacting with the environment. However, the training process is still dominated by performing these interactions. Even for the case with lowest percentage (i.e., NSFNET topology with 64 workers), interacting with the environment still accounted for over 98\% of training time. As a result, we could increase the number of workers even further than we did in our experiments.

\subsection{Training speed-up relative to PPO}

In this experiment we compare the training time of the ES method against the traditional PPO with backpropagation. First, we measure the reward at which PPO converges. This is done by running PPO with a redundant number of iterations, and then take the average reward during the last 50 evaluations. Then, we compare how long does it take for the ES to reach PPO's performance in convergence.

\begin{figure}[t]
    \centering
    \includegraphics[width=0.95\linewidth]{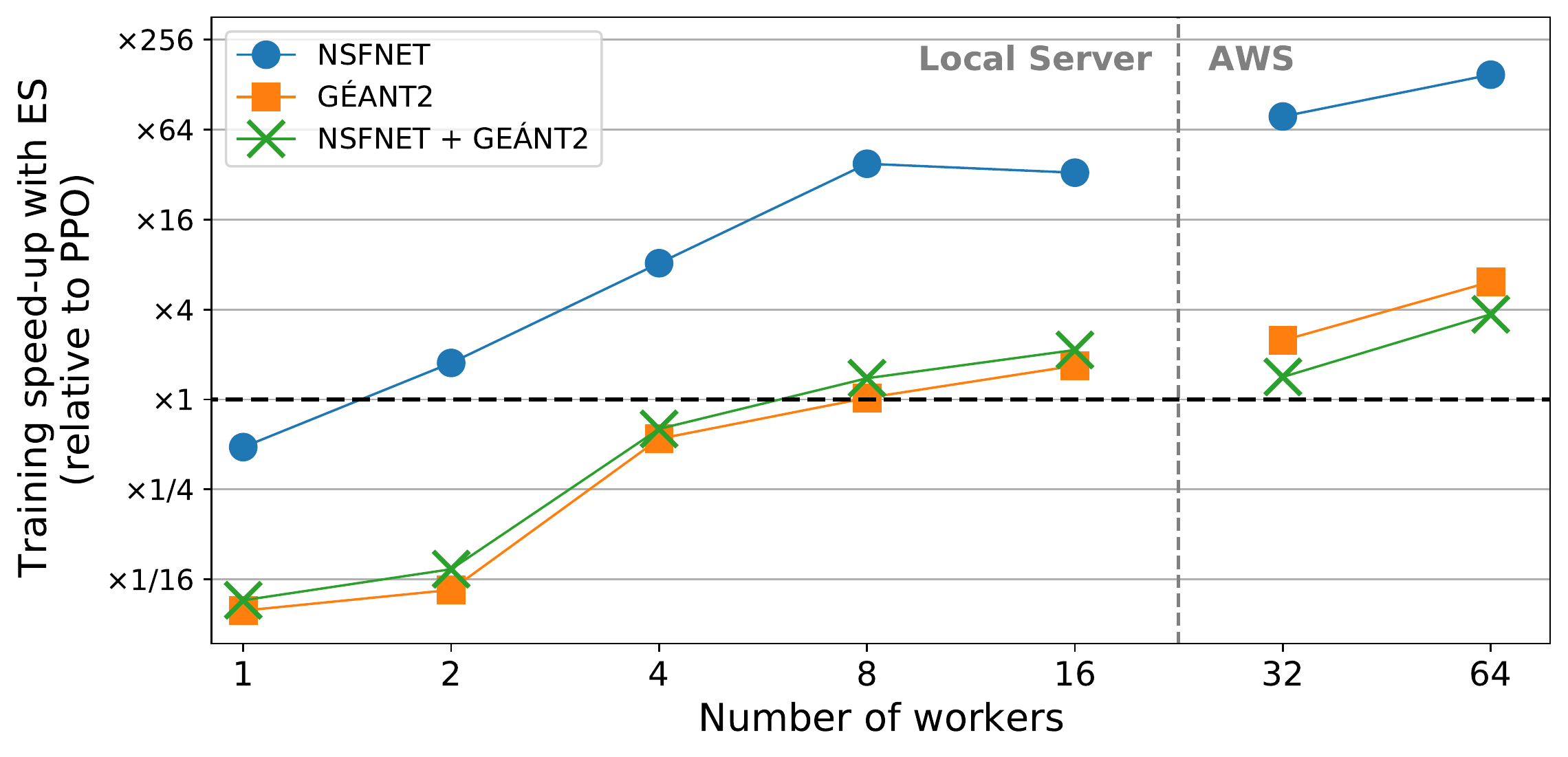}
    \vspace{-2mm}
    \caption{Training speed-up achieved using ES for a different number of workers and topologies. If the evaluation performance is above the value of \textit{x1} in the y-axis, it indicates that ES is faster than PPO. If it is below this value, it indicates that PPO is faster than ES.}
    \label{fig:res_3}
\vspace{-4mm}
\end{figure}

In Figure~\ref{fig:res_3} we can observe the experimental results. The y-axis indicates how much faster is the DRL agent trained with ES than the traditional PPO to converge. The horizontal dotted line at the value of \textit{x1} represents the break-even point where both techniques train just as fast. In other words, if ES's performance is above the line it means that ES is faster than PPO. When the performance is below it means that PPO is faster than ES. For example, when using 2 workers to train a DRL agent on NSFNET, ES are almost two times faster than the traditional PPO.

The general trend indicates that a higher number of workers leads to a better performance relative to PPO. However, there is a clear difference between the models being trained on the NSFNET topology and on the GEANT2 topology. The results from the NSFNET topology indicate that ES surpasses PPO with only 2 workers. The speed-up only stalls when it reaches 16 workers. This can be attributed to ES's reliance in randomness to update the solution. As a result, the model trained with 16 workers required more iterations before it reached PPO's performance due to the poor quality of the perturbations. When considering the executions in the remote server we see that increasing the number of workers still results in an even more accelerated training.

The evaluation results on the GEANT2 topology and on the mixed scenario with GEANT2 and NSFNET show a more modest speed-up. Specifically, the results indicate that more workers are needed to surpass PPO than in the case of training only on NSFNET. The performance gap between these scenarios highlights the weakness of ES: larger network optimization scenarios involve an increased training time for ES methods. This is because the GNN becomes larger, resulting in an increased number of mutations needed to try in each iteration.

\section{Discussion}

Training methods that use ES have several advantages compared to other traditional DRL techniques. First and utmost, it is trivial to parallelize training in ES. This is because the task of evaluating the mutations, the most expensive part of the algorithm, can be split across multiple workers. ES also has less hyperparameters to fine-tune, being the number of mutations the most significant one. This results in a faster experiment setup. Additionally, ES benefits in avoiding several pitfalls of DRL. For example, ES do not have issues with environments with sparse rewards or they do not need for future rewards discount\cite{salimans2017evolution}.

However, ES have two major disadvantages when compared to DRL. First, as it learns from sets of episodes rather than individual interactions, it also means that ES take a longer effort to update its solution. Also, ES cannot be executed in environments with potentially infinite episodes, as ES requires for the episode to end to evaluate its mutations. In practice a maximum episode length can be forced, but that means that the agent is not guaranteed to perform correctly after that point. 

\section{Related work}

The first initial attempts to accelerate DRL's training was with A3C \cite{mnih2016asynchronous}, in which several workers combined efforts to train a single centralized network. Shortly thereafter it was improved into A2C \cite{clemente2017efficient}, a synchronous version of A3C which proved to perform better. There was also GA3C \cite{babaeizadeh2016reinforcement}, which consisted in adapting A3C for it to take advantage of GPUs.

From the most recent related work, Ape-X \cite{horgan2018distributed}, D4PG \cite{barth2018distributed} and IMPALA \cite{espeholt2018impala} are among the most widely used methods to accelerate DRL. While Ape-X is based on DQN and the others on actor-critic models, all three share the same paradigm: to accelerate interactions by having multiple actors interacting with the environment, all feeding a common experiences buffer so a learner can train the model.

This is in contrast to the use of ES, which replaces the need for a buffer altogether and it is not based on any of the original RL models, instead opting for an approach using evolutionary algorithms for optimization.

\section{Conclusion}

In this paper we explored the use of ES to accelerate DRL agents for network optimization. The experimental results showed that increasing the number the workers resulted in a linear decrease in the training time of the DRL agent. Specifically, the results showed that we achieved over a 128-fold speed-up for the NSFNET topology, and a 6-fold speed-up for the GÉANT2 topology. On top of that, ES provides additional advantages, such as requiring fewer hyperparameters and mitigating problems with RL algorithms, such as the ones associated with sparse rewards and non-MDP environments.

As a result, ES should be considered as a viable alternative to accelerate DRL algorithms. However, results also suggest that ES does not scale to larger problems as we initially hoped. As a result, choosing which approach may be better to accelerate the DRL training may depend on the nature of the environment, as well as on the number of trainable parameters of the DRL agent.

\section*{Acknowledgment}

This publication is part of the Spanish I+D+i project TRAINER-A (ref.PID2020-118011GB-C21), funded by MCIN/ AEI/10.13039/501100011033. This work is also partially funded by the Catalan Institution for Research and Advanced Studies (ICREA) and the Secretariat for Universities and Research of the Ministry of Business and Knowledge of the Government of Catalonia and the European Social Fund.

\bibliographystyle{IEEEtran}
\bibliography{references}

\end{document}